\documentclass[conference]{IEEEtran}
\usepackage{textcomp}
\usepackage{xcolor}
\usepackage{graphicx}
\usepackage{graphics}
\usepackage{subcaption}
\usepackage{balance}
\usepackage{comment}
\usepackage{amsmath}
\usepackage[ruled,linesnumbered]{algorithm2e}
\usepackage{multirow}
\usepackage{makecell}
\usepackage{float}
\usepackage{hyperref}

\usepackage{subcaption}
\begin{document}
\title{Exploring the Sparsity-Quantization Interplay on a Novel Hybrid SNN Event-Driven Architecture}

\author{Ilkin Aliyev, Jesus Lopez, and Tosiron Adegbija\\
Department of Electrical and Computer Engineering\\
The University of Arizona, Tucson, AZ, USA\\
Email: \{ilkina, jlopezramos, tosiron\}@arizona.edu

}

\maketitle

\begin{abstract}

Spiking Neural Networks (SNNs) offer potential advantages in energy efficiency but currently trail Artificial Neural Networks (ANNs) in versatility, largely due to challenges in efficient input encoding. Recent work shows that direct coding achieves superior accuracy with fewer timesteps than traditional rate coding. However, there is a lack of specialized hardware to fully exploit the potential of direct-coded SNNs, especially their mix of dense and sparse layers. This work proposes the first hybrid inference architecture for direct-coded SNNs. The proposed hardware architecture comprises a dense core to efficiently process the input layer and sparse cores optimized for event-driven spiking convolutions. Furthermore, for the first time, we investigate and quantify the quantization effect on sparsity. Our experiments on two variations of the VGG9 network and implemented on a Xilinx Virtex UltraScale+ FPGA (Field-Programmable Gate Array) reveal two novel findings. Firstly, quantization increases the network sparsity by up to $15.2\%$ with minimal loss of accuracy. Combined with the inherent low power benefits, this leads to a $3.4\times$ improvement in energy compared to the full-precision version. Secondly, direct coding outperforms rate coding, achieving a $10\%$ improvement in accuracy and consuming $26.4\times$ less energy per image. Overall, our accelerator achieves ~$51\times$ higher throughput and consumes half the power compared to previous work. Our accelerator code is available at: \url{https://github.com/githubofaliyev/SNN-DSE/tree/DATE25}.

\end{abstract}


\section{Introduction}
Spiking Neural Networks (SNNs) offer a biologically inspired, energy-efficient alternative to traditional Artificial Neural Networks (ANNs) by utilizing sparse, event-driven computations. SNNs mimic the brain's neurons by communicating through discrete pulses or "spikes". However, achieving accuracy comparable to ANNs, especially in deep SNNs, is challenging due to the complexities of encoding input data into spikes \cite{guo2021neural, rueckauer2018conversion}. Direct coding \cite{wu2019direct} is a promising approach that addresses this challenge by repeatedly presenting input samples over multiple timesteps, improving SNN accuracy with fewer timesteps compared to the popular rate coding approach \cite{kim2022rate}. In direct coding, the raw, floating-point input data is directly fed into the first convolution layer, which generates floating-point outputs. These floating-point outputs are processed by a spiking neuron layer that implements a threshold-based spike generation mechanism to produce the binary spikes that drive the rest of the SNN.

While prior work suggests that rate-coded networks may be more energy efficient due to the non-binary activations in direct-coded SNN input layers \cite{kim2022rate}, we argue that the true energy-saving potential of direct coding remains largely untapped. Existing evaluations \cite{kim2022rate} are constrained by fixed timestep comparisons and platforms optimized for dense computations, overlooking the inherent benefits of variable sparse activity patterns characteristic of direct-coded SNNs. Moreover, traditional event-driven \cite{sommer2022efficient} or systolic array-based SNN implementations \cite{aung2023deepfire2}, while efficient for homogeneous computations, struggle to effectively handle the diverse layer characteristics present in direct-coded SNNs. This mismatch often leads to underutilization of processing elements in layers requiring fewer resources, further diminishing overall efficiency. To fully realize the potential of direct-coded SNNs, heterogeneous architectures with specialized hardware for both dense and sparse layers are essential.


In this paper, we propose the first-of-its-kind hybrid architecture explicitly designed for direct-coded SNNs. Our architecture leverages workload partitioning, assigning the input layer---characterized by the largest feature map dimensions, non-binary, and non-sparse activations---to a dedicated dense core. The remaining network layers, exhibiting varying degrees of sparsity, are processed by specialized sparse cores. Our hardware design is further distinguished by novel, parameterized, fully-pipelined, and high-throughput datapaths. We establish fine-grained, \textit{layer-wise} workload models that encompass individual layer sizes, input feature map dimensions, input activation sparsity, and other key factors. These models serve as the foundation for design-time parameter selection, enabling the optimal partitioning of heterogeneous hardware resources to maximize performance. Furthermore, we conduct the first ablation study to investigate the impact of quantization on the intrinsic sparsity behavior of direct-coded SNN models, an important step toward realizing energy-efficient SNN implementations. 

Our results on a Xilinx Virtex UltraScale+ FPGA across multiple datasets demonstrate the significant advantages of our hybrid architecture, quantization, and direct coding for SNNs. Quantization (with 4-bit integer weights and biases) substantially reduces spiking traffic, while direct coding achieves state-of-the-art accuracy with minimal timesteps, countering prior findings \cite{kim2022rate} and highlighting the potential for reduced hardware latency and improved energy efficiency. These results underscore the critical need for specialized architectures to fully unlock the efficiency potential of SNNs. Notably, compared to the most related work \cite{gerlinghoff2022resource}, our accelerator achieves $50\%$ lower power and $51\times$ higher throughput for a spiking VGG9 implementation on CIFAR100.

\begin{figure}[t!]
\vspace{-5pt}
		\centering
		\includegraphics[width=0.55\linewidth]{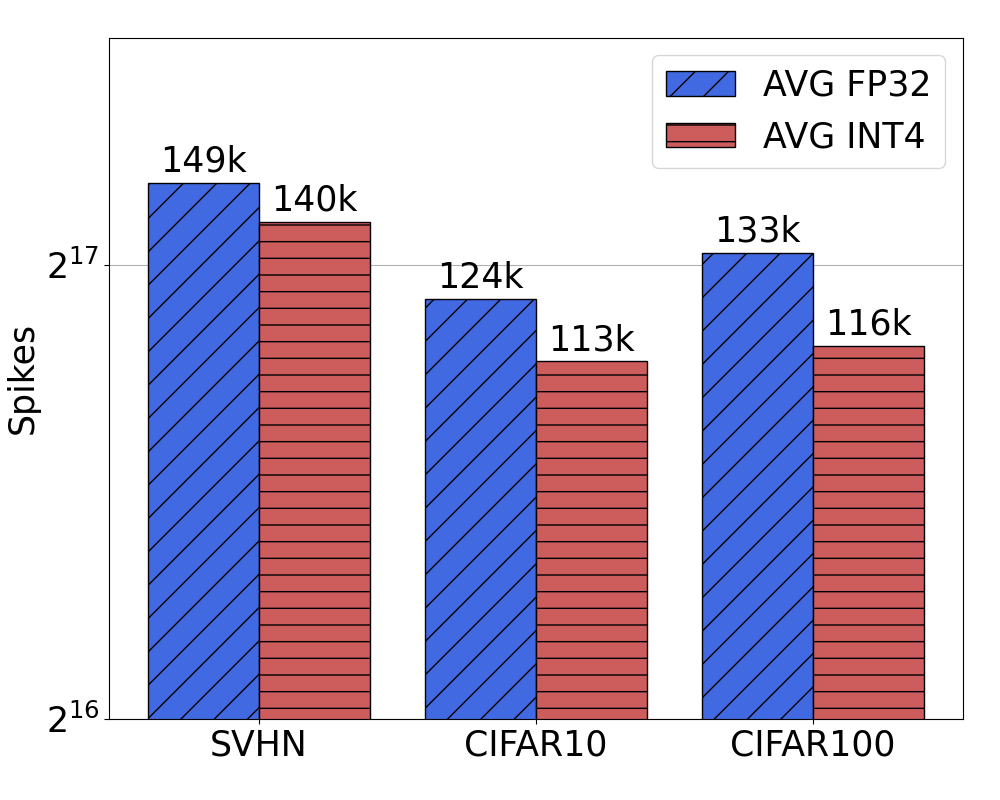}
		\vspace{-8pt}
		\caption{Quantization effect on the total number of spikes across 3 different datasets. The model's accuracy for \texttt{fp32} vs \texttt{int4} is ($94.3\%$, $93.8\%$), ($86.6\%$, $86.2\%$) and ($57.3\%$, $54.2\%$) for SVHN, CIFAR10 and CIFAR100 respectively. The \texttt{int4} version of the datasets results with $6.1\%$, $10.1\%$, and $15.2\%$ fewer spikes than the \texttt{fp32}.}
		\label{fig:quant_sprst}
\end{figure}


\section{Background} \label{sec:background}
\subsection{Spiking Neuron Model}
The spiking neuron model used in this work is a leaky integrate-and-fire (LIF) neuron \cite{lansky2006parameters}, whose characteristics are shown in Equations \ref{ref:lif1} and \ref{ref:lif2}.

\begin{equation} \label{ref:lif1}
    u_{j}[t+1] = \beta \cdot u_{j}[t] + \sum_{i} w_{ij} \cdot s_{i}[t] - s_{j}[t] \cdot \theta
\end{equation}

\begin{equation}\label{ref:lif2}
s_{j}[t] = 
\begin{cases} 
1, & \text{if } u_{j}[t] > \theta \\
0, & \text{otherwise}
\end{cases}
\end{equation}

Equation \ref{ref:lif1} describes how the neuron's membrane potential ($u_{j}[t]$) evolves over time. The decay factor ($\beta$), ranging between $0$ and $1$, controls how much the previous potential $u_{j}[t]$ affects the current potential $u_{j}[t+1]$. A higher $\beta$ value implies less decay, enabling the neuron to retain more of its previous state, resulting in sparser behavior. Incoming weighted spikes from other neurons ($w_{ij} \cdot s_{i}[t]$) increase the potential, while a threshold-based self-decay term ($s_{j}[t] \cdot \theta$) reduces it. Equation \ref{ref:lif2} determines when the neuron fires. If the membrane potential exceeds the threshold ($\theta$), the neuron generates a spike ($s_{j}[t] = 1$). As such a lower $\theta$ value reduces the potential required for firing, thereby increasing the neuron's firing frequency.

\begin{figure*}[!t]
		\centering
		\includegraphics[width=0.95\linewidth]{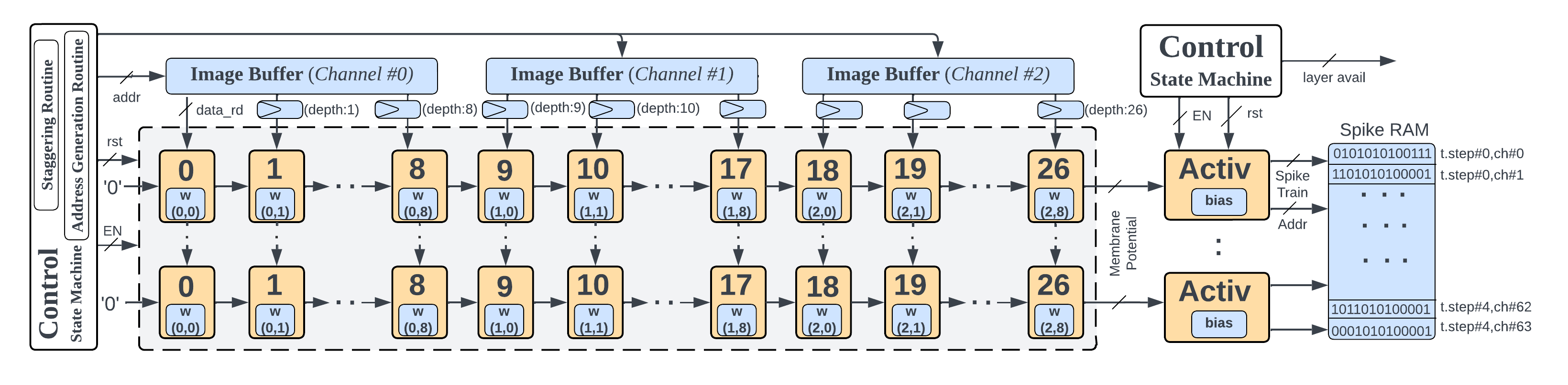}
		\vspace{-8pt}
		\caption{Dense Core (DC) hardware architecture. Weight stationary dataflow - PEs in a row collectively work on one output channel.}
		\label{fig:dc}
		\vspace{-15pt}
\end{figure*}

\subsection{Quantization-Aware Training}
We employ \textit{Quantization-Aware Training (QAT)} \cite{jacob2018quantization} to quantize model weights and biases into integers, with the quantization error incorporated into the loss function during training. This enables the model to adapt to the quantization constraints, enhancing its performance when operating with quantized data. During evaluation, weights and biases are fully quantized, while neuronal parameters remain in floating-point due to the current lack of support for LIF neuron quantization. The accumulated membrane data undergoes dequantization back to floating-point for accurate spiking operations.

\section{Impacts of Quantization on Sparsity}
The potential efficiency gains of low-bit quantization raise an intriguing question: \textit{can quantization influence the sparsity of an SNN model?} To investigate this, we empirically compared a 32-bit floating-point network (called \texttt{fp32}) with its 4-bit integer (\texttt{int4}) counterpart, both trained using \textit{QAT}. Section \ref{sec:results} details our experimental setup. Figure \ref{fig:quant_sprst} illustrates the variability in the VGG9 model's sparsity for SVHN, CIFAR10, and CIFAR100. Our experiments revealed two key insights: 

\begin{enumerate}
    \item \textit{Comparable accuracy with both precisions:} The \texttt{int4} and \texttt{fp32} networks demonstrate similar accuracy varying by only $0.5\%, 0.4\%,$ and $3.1\%$ for SVHN, CIFAR10, and CIFAR100, respectively.
    \item \textit{Quantization increases sparsity:} Compared to \texttt{fp32}, \texttt{int4} yields $6.1\%$, $10.1\%$, and $15.2\%$ fewer spikes for SVHN, CIFAR10, and CIFAR100, respectively. We hypothesize that introducing quantization noise during training implicitly encourages sparsity by selectively deactivating neurons, leading to a sparser and more efficient representation without sacrificing accuracy. 
\end{enumerate} 

These findings suggest that quantization, beyond its traditional role in reducing model size and power consumption, may also induce sparsity. This opens exciting avenues for further exploring the synergy between quantization and sparsity-aware techniques to further enhance SNN efficiency.

\section{Hybrid Hardware Architecture} \label{sec:arch}
\noindent\textbf{Architecture Overview}: Our hybrid architecture integrates specialized dense and sparse cores connected by on-chip FPGA memories.
Core allocation follows a \textit{layer-wise} strategy, with each core's size tailored to the corresponding layer's sparsity-driven workload needs. This \textit{layer-wise} resource specialization is essential due to the potential for high sparsity variability across network layers. While our evaluation focuses on specific model architectures, the principles underlying this hybrid design---specialized processing cores and layer-wise resource allocation---can be readily adapted to support various SNN models and computational requirements.

\subsection{Dense Core}
Figure \ref{fig:dc} depicts the block diagram of the dense core (dc), comprising three key components: a \textit{PE array}, \textit{control} units, and \textit{activation} units. The \textit{PE array} features a fixed column of 27 processing elements (PEs) and employs a \texttt{weight stationary (\textit{WS})} dataflow. We chose 27 PEs because of the \textit{WS} requirements, specifically to handle $3$ input channels and $3\times3$ filters. This choice minimizes memory footprint requirements, as the input layer's three channels lead to fewer weights compared to input or output activations. Since we employ tiling in the output channel, we parameterize the number of rows, meaning that each row handles one output feature map. Each row then sequentially moves onto the next feature map until all output channels are processed. As a result, partial sums flow horizontally (left to right) while input image pixels flow vertically (top to down) in a systolic manner.

Each PE contains a Multiply and Accumulate (MAC) unit, along with a register for weight storage. All 27 PEs work in parallel to process input channels and generate a single output membrane potential per cycle. To maintain the systolic pipeline flow, we utilize independent registers/FFs to act as ``shift registers" to provide a fixed delay for each of the 27 inputs in the array. These registers are controlled by the \textit{Staggering Routine} within the \textit{Control} unit. The right-most PE's shift register depth dictates the array's pipeline depth.

The \textit{Control} unit is implemented as a state machine and orchestrates three main tasks:

\begin{itemize}
    \item \textit{Data management:} It feeds image data into the \textit{PE array}. The \textit{Address Generation Routine} calculates read indices for the 27 top PEs, accessing image buffers (implemented with on-chip flip-flops for their small storage footprint). This enables row-major image storage while facilitating parallel weight access by all 27 PEs. The \textit{Staggering Routine} manages "shift registers" that control the data flow from the image buffers.
    \item \textit{Activation synchronization:} It starts/stops the \textit{Activ} unit using the \textit{EN} signal. Once the PE pipeline is filled (i.e., the first membrane potential accumulation is produced), the \textit{Control} unit triggers the activation phase (responsible for spike train generation) with the signal \textit{EN}. 
    \item \textit{Output channel tiling:} The \textit{Control} unit coordinates the tiling of output channels, meaning that each row in the systolic array strides through the output channels. When the rows of the systolic array transition to the next output feature map computation, it resets the partial sums and membrane potentials within the \textit{PE Array} and \textit{Activ} units using the \textit{rst} signal. Finally, the unit sets the \textit{layer avail} signal high to indicate the completion of layer processing.
\end{itemize}

The accumulated potential from each row flows into the \textit{Activ} unit, where it performs two core functions. First, it modifies the incoming potential by adding the filter bias value, applying the leakage factor ($\beta$) to mimic the leaky behavior, and then performs thresholding. If the membrane potential exceeds the threshold ($\theta$), the unit subtracts the threshold value from the membrane potential and sets the associated spike to $1$; otherwise, the spike is $0$. Secondly, after processing an entire output feature map, the \textit{Activ} unit writes the generated spike train to the Block RAM (BRAM), which serves as the intermediate storage between layers. Spike trains in BRAM follow a timestep-major order (see Figure \ref{fig:dc}) with the spike trains for consecutive timesteps stored in contiguous addresses. For example, if the layer consists of $N$ output channels and $T$ timesteps, then $N \times T$ locations in total are required for the layer's spike train storage. This layout is consistent for both dense and sparse cores.

\subsection{Sparse Core}
\begin{figure}[!t]
		\centering
		\includegraphics[width=0.8\linewidth]{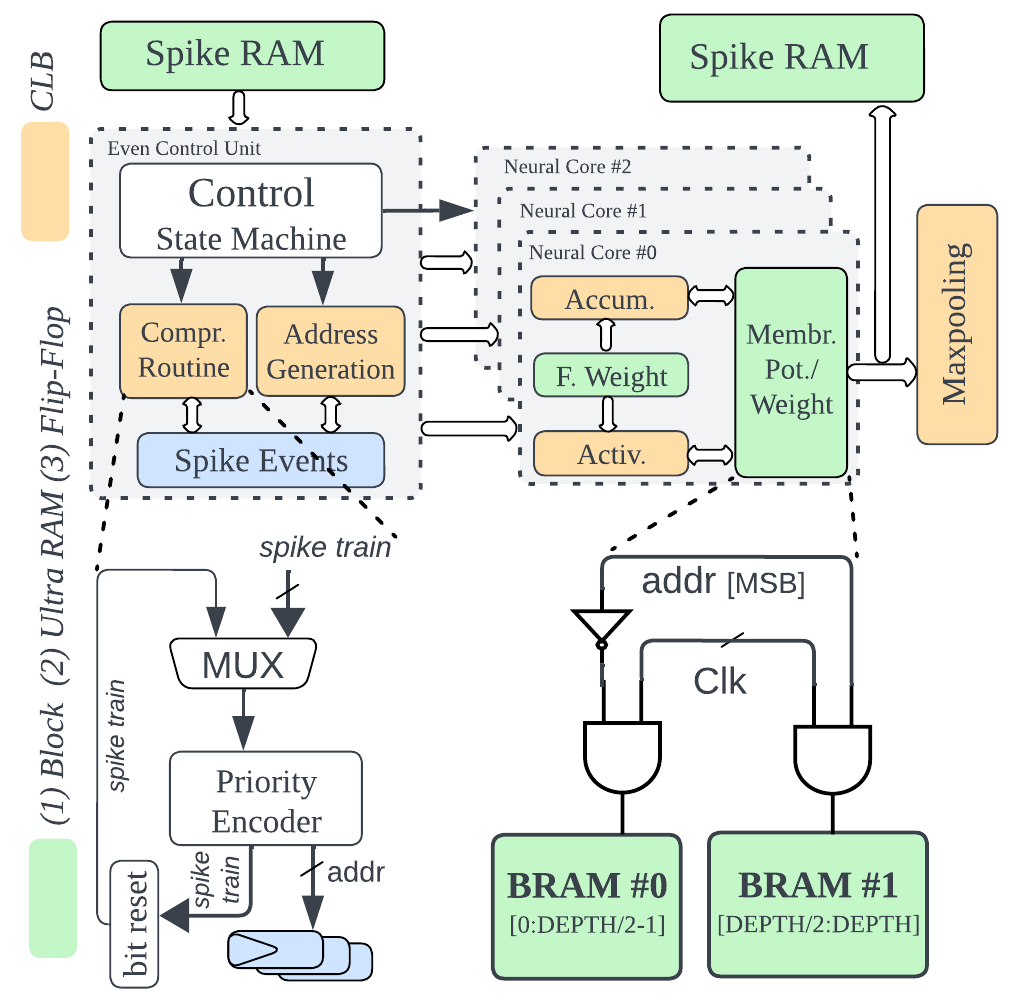}
		\vspace{-8pt}
		\caption{Sparse Core (SC) hardware architecture.}
		\label{fig:sc}
\end{figure}

We adapt conventional convolution (CONV) and fully connected (FC) layers for  \texttt{event-driven} processing by dividing their operation into spike train compression and accumulation phases. The sparse core design, illustrated in Figure \ref{fig:sc}, implements these phases and consists of two primary components: an \textit{Event Control Unit (ECU)} and \textit{Neural Cores (NC)}. Within the ECU, the \textit{control} routine is implemented as a state machine and has several key functions. 

\par\noindent\textbf{Compr. Routine:} First, the ECU fetches a spike train from the input \textit{Spike RAM}, and then tiles the whole spike train into $n$-bit chunks, which are processed sequentially by the Compression (\textit{Compr.}) routine. The core function of this routine is to eliminate the non-spiking (i.e., '0') bits from the bit array and generate a compact register array called $Spike Events$ (see Figure \ref{fig:sc}). This is achieved by processing $n$ bits per cycle and using a priority encoder to identify and send the address of the first set bit ('1') within each chunk to the $Spike Events$ array, a common application of priority encoders, as shown in Figure \ref{fig:sc} (right). Finally, the ECU's \textit{bit reset} component sets the identified '1' bit back to '0' in the previous cycle's spike train, allowing the \textit{Compr.} routine to locate the next set bit in subsequent cycles.

\par\noindent\textbf{Address Generation:} After compression, the accumulation phase begins, utilizing the $Spike Events$ register array. For each spike, there are 9 associated neurons in the feature map whose membrane potentials must be updated with corresponding filter coefficients. The \textit{Address Generation} routine iterates through the filter coefficients and calculates the $(row,col)$ addresses of neurons associated with that spike event and filter size (e.g., a spike at $(row,col)$ affects the 9 neurons from $(row-3,col-3)$ to $(row,col)$). These $(row,col)$ signals are then directed to the neural cores (\textit{NCs}). Importantly, the \textit{Compr.} and accumulation phases execute in parallel, enabling the compression of new spike trains from subsequent input feature maps while simultaneously processing earlier feature map spike events. 

With the $(row,col)$ pairs provided, the \textit{Accum} routine within each \textit{NC} performs the accumulation phase of the LIF neuron. It reads the membrane potential value from the BRAM, updates them with corresponding coefficient weights, and writes back the result to the BRAM. Similar to the dense core, the sparse cores also support quantization. Also, note that both the \textit{Address Generation} and \textit{Accum} routines are fully pipelined and can update one neuron per cycle. Filter weights for CONV layers are stored in BRAM and LUTRAM, while larger fully-connected (FC) weights use Ultra RAMs (URAMs) for their higher density and energy efficiency. After processing all input feature maps, the \textit{control} unit enables the \textit{NC'}s \textit{Activ.} routine to initiate the LIF neuron's spiking phase.

The architecture unrolls the output channels by a factor of $N$, defined as a top-level parameter, to determine the number of \textit{NC} instances. Each \textit{NC} instance strides through the output channels by $N$. For instance, an \textit{NC} instance with index $2$ and $N=8$ will process OFMs with indices 1, 9, 17, etc. The architecture also features max-pooling on spikes, which aligns better with SNN temporal dynamics and minimizes information loss during downsampling, compared to other pooling methods \cite{fang2021incorporating}. Implementation on binary feature maps only requires sliding an OR gate over an $N\times N$ input area, where $N$ is the downsampling ratio.



\subsection{Memory Optimization} \label{sec:mem_opt}
For efficient resource usage, we only use the FPGA's on-chip memory (BRAM, URAM, LUTRAM) for storing model parameters and spike trains, avoiding the external DDR memory. Early layers with smaller convolutions use LUTRAM, which is more efficient than flip-flops for small RAMs due to its flexibility, despite being less abundant than flip-flops in Xilinx UltraScale+ devices. 

\begin{table}[t!]
\caption{Area utilization and power estimate comparison of the proposed hardware tailored for CIFAR100.}
\label{table:base_area_power}
\centering
\begin{tabular}{c | c | c | c } 
 \hline
 \multirow{2}{*}{\textbf{Layer}} & \multirow{2}{*}{\textbf{LUT \& FF}} & \multirow{2}{*}{\textbf{BRAM \& URAM}} & \textbf{Power*} \\ [0.5ex] 
 &&&[W] \\
 \hline \hline
\multicolumn{4}{|c|}{\texttt{int4} hardware} \\
 \hline \hline
 \textit{CONV\_1\_1} & 1.9  \& 1.9K    & 0 \& 0 & 0.048 \\ 
 \textit{CONV\_1\_2} & 11.7K  \& 14.6K & 32 \& 0 & 0.205  \\
 \textit{CONV\_2\_1} & 1.7K   \& 2.1K  & 44 \& 0 & 0.054  \\
 \textit{CONV\_2\_2} & 5.1K   \& 5.1K  & 164 \& 0 & 0.17  \\
 \textit{CONV\_3\_1} & 1.6K   \& 1.3K  & 144 \& 0 & 0.1 \\ 
 \textit{CONV\_3\_2} & 5.7K   \& 5.2K  & 216 \& 0 & 0.293 \\
 \textit{CONV\_3\_3} & 5.8K   \& 5.1K  & 211 \& 0 & 0.284 \\
 \textit{FC}         & 6K     \& 2.1K  & 168 \& 0 & 0.125  \\
 \hline
 Total & 109.7K \& 37.6K & 979 \& 0 & 1.231 \\
  \hline
 Utilization & 6.43\% \& 1.11\% & 30.23\% \& 0 & -- \\
 \hline \hline  
 \multicolumn{4}{|c|}{\texttt{fp32} hardware} \\
 \hline \hline
 \textit{CONV\_1\_1} & 11.6K \&  1.9K & 0 \& 0 & 0.051 \\ 
 \textit{CONV\_1\_2} & 670.3K \& 15.2K & 32 \& 0 & 0.251  \\
 \textit{CONV\_2\_1} & 11.4K \&  5.3K & 212 \& 0 & 0.152  \\
 \textit{CONV\_2\_2} & 34.4K \& 10.1K & 272 \& 54 & 0.561  \\
 \textit{CONV\_3\_1} & 11.6K \&  2.9K & 464 \& 129 & 0.405 \\ 
 \textit{CONV\_3\_2} & 45.6K \& 12.5K & 648 \& 145 & 0.96 \\
 \textit{CONV\_3\_3} & 39.2K \&  8.4K & 631 \& 140 & 0.634 \\
 \textit{FC}         & 7.6K \& 2.8K  & 607 \& 368& 0.508 \\
 \hline
 Total            &  821.6K \& 58.7K & 2466 \& 836 & 3.471 \\
 \hline
 Utilization & 47.73\% \& 1.73\% & 91.2\% \& 65.69\% & -- \\
 \hline
\end{tabular}
\textit{*Instance-level dynamic power. Static power is $3.13W$ and $3.22W$ for \texttt{int4} and \texttt{fp32} respectively.}
\end{table}

Our strategy for aggressive on-chip storage, while beneficial, comes with a trade-off in power consumption. Since a large portion of this storage remains inactive during layer execution (until \textit{NC} reaches the next output channel), the associated weight data is not used. To address this, we implement clock gating for our memory units (right side of Figure \ref{fig:sc}). The memory is partitioned into two regions, with the most significant bit (MSB) of the address line controlling the active region for both reads and writes. An \texttt{AND} gate controls the clock signal to the memory unit, ensuring that only the active region receives clock cycles. While this on-chip storage strategy works well for quantized models like \textit{int4}, it poses limitations for \textit{fp32} models. Specifically, the high LUT usage for weight storage prevents us from scaling beyond the custom VGG9 architecture or deploying larger models. 

\begin{figure*}[t]
  \centering
  \subfloat[SVHN]{\includegraphics[width=0.29\textwidth]{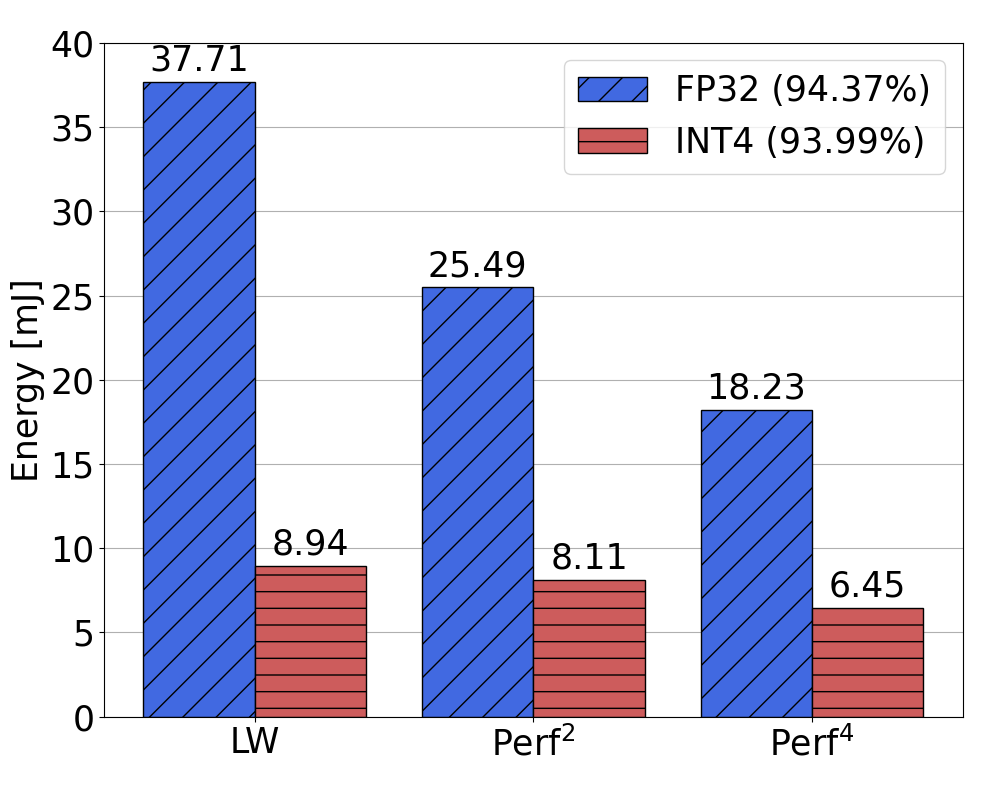}\label{fig:svhn}}
  \subfloat[CIFAR10]{\includegraphics[width=0.29\textwidth]{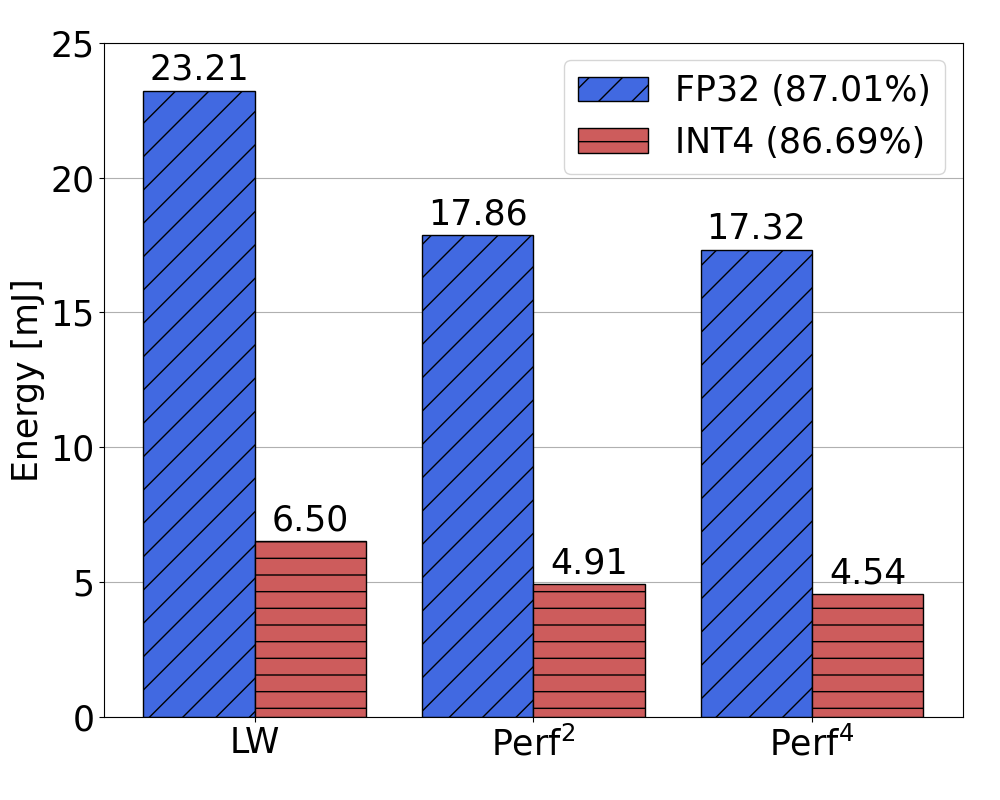}\label{fig:cifar10}}
  \subfloat[CIFAR100]{\includegraphics[width=0.29\textwidth]{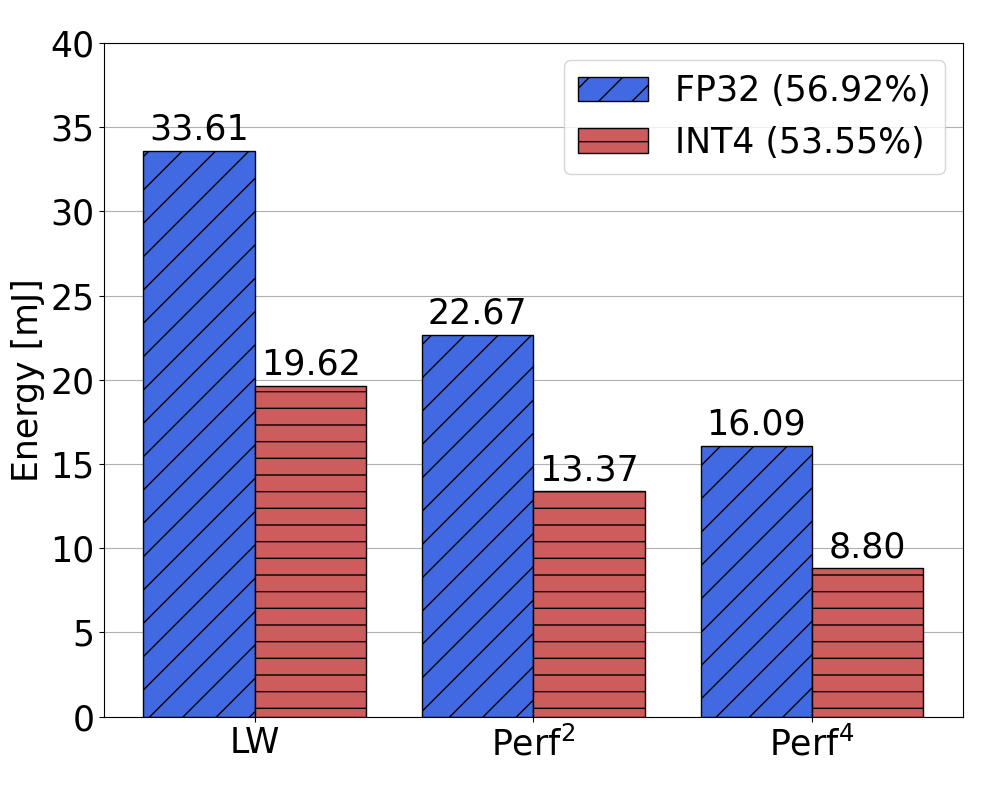}\label{fig:cifar100}}
  \vspace{-5pt}
  \caption{Energy comparison for \texttt{fp32} vs \texttt{int4} hardware. The \texttt{LW} configurations are $({1, 7, 1, 8, 2, 4, 14, 1, 2})$, $(1, 8, 4, 18, 6, 6, 20, 2, 1)$ and $(1, 7, 3, 12, 4, 18, 16, 4, 1)$ for SVHN, CIFAR10 and CIFAR100 respectively}
      \label{fig:quant_energy}
    \vspace{-10pt}
\end{figure*}

\subsection{Quantization Support} 
Due to the lack of native quantization support, the design incorporates floating-point computation capabilities in both dense and sparse cores to de-quantize weights and biases retrieved from memory. We utilize a resource-efficient \texttt{shift-and-add} approach instead of DSP blocks to perform multiplications by constants \cite{dinechin2019table}, improving overall efficiency.  

\section{Experimental Results} \label{sec:results}
In this section, we first detail our experimental setup and leverage our novel hybrid architecture to analyze the benefits of quantization (resource utilization, energy, power, and latency). We then compare direct and rate coding, demonstrating the superiority of direct coding. Finally, we compare our hardware against two recent and relevant works.

\subsection{Experimental Setup} 
We implemented our proposed hardware using SystemVerilog and synthesized on a Xilinx Virtex\textregistered{} UltraScale+\texttrademark{} XCVU13P FPGA. For evaluation, we use the CIFAR-10, CIFAR-100, and Street View House Numbers (SVHN) datasets with a modified VGG9 network trained using \textit{snnTorch} \cite{eshraghian2023training}. The network, trained using surrogate gradients \cite{neftci2019surrogate}, has a structure: 64C3-112C3-MP2-192C3-216C3-MP2-480C3-504C3-560C3-MP2-1064-$P$. $X$C$Y$ denotes $X$ filters of size $Y\times Y$ and MP$Z$ denotes $Z \times Z$ maxpooling. $P$ represents the number of neurons (i.e., population) in the network's output layer. Prior work \cite{aliyev2023design} has shown that using a population of neurons in the output layer can lead to better accuracy, even with fewer time steps. That is, we can make the output layer larger instead of needing more time steps to achieve good results. Through experimentation, we found that the following $P$ values provided the best accuracy with the minimum possible spike train length: $P = 1000$ for CIFAR10 and SVHN datasets, and $P = 5000$ for CIFAR100. We selected the network architecture to ensure compatibility with the Virtex UltraScale+ platform, following prior work \cite{aung2023deepfire2}. We tuned the \textit{leaky} neuron hyperparameters ($\beta = 0.15$ and $\theta=0.5$) and used  \textit{layer-wise} batch normalization to prevent overfitting.

\noindent\textbf{Hardware configurations:} For each dataset, we have one lightweight (\texttt{LW}) and two performance-optimized configurations (\texttt{perf$^{2}$} and \texttt{perf$^{4}$}). The \texttt{LW} baseline prioritizes minimal resource usage while ensuring balanced \textit{layer-wise} execution latency of the network. The two performance-optimized versions scale up resources by factors of $2\times$ and $4\times$, respectively. To determine the optimal lightweight configuration, we modeled the trained network's \textit{layer-wise} workload distribution using Equation \ref{res_partit}: 

\vspace{-5pt}
\begin{equation} \label{res_partit}
  \begin{aligned}
    W_{\text{CONV}} &= F \times C_{\text{out}} \times \sum_{i=1}^{N} S_i, \quad
    W_{\text{FC}} &= N \times S
  \end{aligned}
\end{equation}

\noindent where \textit{F} is the number of filter coefficients (e.g., $9$ for $3\times3$), \textit{C\_{\text{out}}} is the number of output channels, and $S\_i$ is the number of spikes for input feature map $i$. The spike information was acquired empirically by running the network once on the hardware. Our goal was to partition resources to minimize the execution latency difference between the most and least workload-intensive layers. All three configurations were synthesized to operate at a 100 MHz clock frequency.

\subsection{Resource Utilization}
Table \ref{table:base_area_power} summarizes the implementation results of our proposed hardware for CIFAR100 (in \texttt{perf$^{2}$}), which are representative of the trends observedfor CIFAR10 and SVHN. We compare the results of full-precision (\texttt{fp32}) with its 4-bit integer counterpart (\texttt{int4}). We empirically determined that a $(1, 28, 12, 54, 16, 72, 70, 19, 4)$ configuration (neural cores allocated per layer) yields the most balanced execution profile (layer overheads: 0.9\%, 13.4\%, 13.6\%, 13.8\%, 12.8\%, 12.3\%, 12.9\%, 15.6\%, 4.8\%). The \textit{int4} implementation demonstrates significant resource savings compared to \textit{fp32}, using $8\times$ fewer LUTs, primarily due to efficient LUTRAM usage for the \textit{CONV\_1\_2} weight data. Additionally, \textit{int4} uses $3.4\times$ fewer BRAMs/URAMs due to the minimum 8-bit width configuration of BRAM primitives for \textit{CONV} layers. Overall, the \textit{fp32} and \textit{int4} designs occupy $24\%$ and $34\%$ of the FPGA's LUT resources, respectively. Importantly, the \texttt{fp32} hardware consumes $2.82\times$ more power than the \texttt{int4} configuration, highlighting the power efficiency benefits of quantization.

\begin{table}[b]
\centering
\footnotesize
\caption{Direct vs. rate coding using CIFAR10. \textbf{Imprv.} represents the energy improvement in direct vs. rate coding}
\begin{tabular}{| c | c | c | c | c | c | c |} 
 \hline
 \multirow{2}{*}{\textbf{Coding}} & \textbf{Time} & \textbf{Total} & \textbf{Acc.} & \textbf{Latency} & \textbf{Energy} & \textbf{Energy}\\  
 &\textbf{Steps}&\textbf{Spikes}&[\%]&[ms]&[mJ]&{\textbf{Imprv.}} \\
  \hline
  \hline
 \multirow{2}{*}{Rate} & \multirow{2}{*}{25} & \multirow{2}{*}{107K} & \multirow{2}{*}{77.37} & \multirow{2}{*}{340} & \multirow{2}{*}{201} & \multirow{2}{*}{---} \\
 &&&&&&\\
   \hline
 \multirow{2}{*}{Direct} & \multirow{2}{*}{2} & \multirow{2}{*}{41K} & \multirow{2}{*}{87.01} & \multirow{2}{*}{11.7} & \multirow{2}{*}{7.6} & \multirow{2}{*}{$26.4\times$} \\
 &&&&&&\\ \hline
\end{tabular}
\label{table:encoding}
\end{table}
\begin{table*}[t!]
\caption{Comparison to previous work}
\label{table:prev_work_comparison}
\centering
\begin{tabular}{c | c | c | c | c | c | c | c | c | c | c} 
 \hline
 \multirow{2}{*}{\textbf{Dataset}} & \multirow{2}{*}{\textbf{Study}} & \multirow{2}{*}{\textbf{Network}} & \textbf{Weight} & \textbf{Acc} & \multirow{2}{*}{\textbf{Platform}} & $\textbf{F}_{Max}$ & \textbf{Power} & \textbf{Latency$^{1}$} & \textbf{Energy$^{1}$} & \textbf{Throughput} \\ [0.5ex] 
 &&&\textbf{Precision}&[\%]&&[MHz]&[W]&[ms]&[mJ]&[FPS] \\
 \hline \hline
 \multirow{2}{*}{SVHN}&  \cite{panchapakesan2022syncnn}  & VGG11 & \multirow{2}{*}{4-bit} & 89 & ZCU102 & 200 & 0.4  & ---  & --- & 65 \\ 
 & \textbf{\texttt{perf$^{4}$}} & VGG9 && 93.9 & XCVU13P  & 100 & 0.89  & 61  & 6.4 & 110 \\ 
 \hline
  \multirow{2}{*}{CIFAR10}&  \cite{panchapakesan2022syncnn}  & VGG11 & \multirow{2}{*}{4-bit} & 78 & ZCU102  & 200 & 0.4  & ---  & --- & 62 \\ 
 & \textbf{\texttt{\texttt{perf$^{2}$}}} & VGG9 & & 86.6 & XCVU13P  & 100 & 0.73  & 59  & 4.9 & 120 \\ 
 \hline
  \multirow{2}{*}{CIFAR100}&  \cite{gerlinghoff2022resource}  & VGG11 & \multirow{2}{*}{32-bit} & 60.1 & XCVU13P  & 115 & 4.9  & 210  & --- & 4.7 \\ 
 & \textbf{\texttt{perf$^{4}$}} & VGG9 && 56.9 & XCVU13P  & 100 & 2.35  & 37 & 16.1 & 218 \\ 
 \hline 
\end{tabular}\\
$^{1}$'---' indicates no reported results in prior work.
\vspace{-15pt}
\end{table*}

\subsection{Energy Analysis} 
Our energy analysis comparing \texttt{int4} and \texttt{fp32} (Figure \ref{fig:quant_energy}) reveals significant energy savings benefits of quantization across CIFAR10, CIFAR100, and SVHN datasets. We calculate the energy expenditure per image by summing the energy per layer. For CIFAR10 and CIFAR100, \textit{int4} reduces the average energy by $3.4\times$ and $1.7\times$, respectively, across all configurations. The majority of these savings (roughly $3\times$ for CIFAR10 and $1.5\times$ for CIFAR100) stem directly from \texttt{int4}'s inherent power advantage. Additionally, the \textit{fp32} designs exhibit a higher spike count ($1.1\times$  and $1.15\times$ for CIFAR10 and CIFAR100, respectively), contributing a further $10-15\%$ to the energy difference. For SVHN, the scaling trend differs slightly due to the resource allocation dominance of the \textit{CONV\_1\_2} layer (with more allocated neural cores), leading to a smaller energy gap between \texttt{fp32} and \texttt{int4}. The \texttt{perf$^{4}$} configuration of the quantized hardware achieves a $28\%$ energy reduction compared to its \texttt{LW} counterpart, contrasting with the $52\%$ reduction in the full-precision design. Since the full-precision hardware utilizes LUTRAM for this layer, increasing the \textit{NC} count has a less pronounced impact on power compared to the quantized hardware. Overall, quantization yields a significant $4\times$ reduction in latency and notable energy savings across all datasets. 

\subsection{Direct vs Rate coding comparison} 
Table \ref{table:encoding} depicts our comparison of direct coding and rate coding on CIFAR10 using the quantized \texttt{LW} configuration. 
Because the rate-coded network receives spikes as inputs, it only needs sparse cores, while the direct-coded network needs our hybrid architecture. As such, although both networks run on our hardware architecture, for a fair comparison, we turned off the dense core for the rate-coded network to prevent unnecessary power consumption. The \textit{Total Spikes} represents the spikes across all timesteps. With only 2 timesteps, direct coding yielded much higher sparsity (2.6$\times$ fewer spikes) and achieved $10\%$ higher accuracy than rate coding at 25 timesteps. Further increasing the timesteps plateaued the accuracy for both schemes\footnote{To enable reproducibility of our accuracy and implementation, our training and hardware code are available at \url{https://github.com/githubofaliyev/SNN-DSE/tree/DATE25}}. Importantly, the direct-coded network consumed $26.4\times$ \textit{less} energy than its rate-coded counterpart, contrary to prior work \cite{wu2019direct}.

\subsection{Comparison to previous work}
Table \ref{table:prev_work_comparison} compares our proposed hybrid architecture's dynamic power and throughput to two recent related works: an event-driven design with quantization support \cite{panchapakesan2022syncnn} and a resource-efficient approach \cite{gerlinghoff2022resource}. We focus on dynamic power and throughput as these metrics were reported in prior works. 

Compared to \cite{panchapakesan2022syncnn}, we achieve more than 2$\times$ the throughput for SVHN and CIFAR10, with a power increase of $2.2\times$ and $1.8\times$, respectively (Table \ref{table:prev_work_comparison}). 
In addition to using a lower power board, \cite{panchapakesan2022syncnn} obtained lower power measurements by subtracting the board's baseline power consumption from the accelerator's total runtime power.  Additionally, we outperform their work in inference accuracy ($4\%$ and $8\%$ higher for SVHN and CIFAR10, respectively). Compared to \cite{gerlinghoff2022resource}, our \texttt{perf$^{4}$} configuration on CIFAR100 achieves $51\times$ throughput improvement and $2\times$ power savings, with only a $3.1\%$ accuracy decrease. This work is the most relevant comparison to ours due to the use of similar implementation platforms.


\section{Conclusion} \label{sec:concl}
This paper presented a novel hybrid architecture that strategically combines dense and sparse cores, optimized for the unique workload characteristics of direct-coded SNNs. Our results demonstrate the significant advantages of direct coding, achieving both superior accuracy and hardware performance compared to traditional rate coding. We also highlight the benefits of quantization, showcasing its impact on sparsity and efficiency. The insights from this work lay the foundation for co-design approaches where SNN algorithms and hardware architectures are co-developed to maximize performance and enable more efficient SNN accelerators. 

Future research will further explore direct coding through expanded evaluations with diverse models and datasets, including larger-scale networks that may require external memory access. While our current implementation focuses on models that fit within on-chip BRAM, additional studies are needed to analyze performance impacts when incorporating off-chip memory access for broader model support. This includes investigating techniques to minimize external memory bandwidth requirements and optimize the memory hierarchy for larger networks.


\section*{Acknowledgment}
This work was partially supported by National Science Foundation Grant 1844952. Any opinions, findings, conclusions or recommendations expressed are those of the authors and do not necessarily reflect the views of the NSF.


\end{document}